\documentclass[11pt,a4paper]{article}
\usepackage{jheppub}

\usepackage{graphics}

\usepackage{ulem}

\newcommand{\ba}{\begin{eqnarray}}
\newcommand{\ea}{\end{eqnarray}}

\title{On the quantum stability of Q-balls}

\date{\today}

\author[a]{Anders Tranberg,}

\author[b]{David J. Weir}

\affiliation[a]{Faculty of Science and Technology, University of Stavanger, 4036 Stavanger, Norway}
\affiliation[b]{Department of Physics and Helsinki Institute of Physics, PL 64 (Gustaf H\"allstr\"omin katu 2), FI-00014 University of Helsinki, Finland}

\preprint{HIP-2013-22/TH}

\abstract{ We consider the evolution and decay of Q-balls under the
  influence of quantum fluctuations. We argue that the most important
  effect resulting from these fluctuations is the modification of the
  effective potential in which the Q-ball evolves. This is
  in addition to spontaneous decay into elementary particle
  excitations and fission into smaller Q-balls previously considered
  in the literature, which -- like most tunnelling processes -- are
  likely to be strongly suppressed. We illustrate the effect of
  quantum fluctuations in a particular model $\phi^6$ potential, for
  which we implement the inhomogeneous Hartree approximation to
  quantum dynamics and solve for the evolution of Q-balls in 3+1
  dimensions. We find that the stability range as a function of (field
  space) angular velocity $\omega$ is modified significantly compared
  to the classical case, so that small-$\omega$ Q-balls are less
  stable than in the classical limit, and large-$\omega$ Q-balls are
  more stable.  This can be understood qualitatively in a simple way.
}


\begin{document}

\maketitle

\section{Introduction}
\label{sec:intro}

Q-balls are a type of localised, periodic solution to certain
non-linear field theories, with a net electric
charge~\cite{Qballs}. They are a form of nontopological soliton~\cite{Friedberg:1976me}. When they exist, Q-balls are classically stable under small perturbations of their shape and radial field profile. In particular, they are spherically symmetric, and the profile can generically be written as (for a complex scalar field $\phi$)
\ba
\label{eq:ansatz}
\phi(t,r) = \frac{\sigma_{\omega}(r)}{\sqrt{2}} e^{i\omega t},
\ea
with $\sigma_{\omega}(r)$ some $\omega$-dependent, but constant in time, profile function. Then the charge is
\ba
Q=\int d^3x\,i \left[(\partial_t\phi)^\dagger\phi-\phi^\dagger(\partial_t\phi)\right] =8\pi\omega \int  r^2\, \sigma^2_{\omega}(r)dr.
\ea
The profile function $\sigma_\omega(r)$ can be found by numerically solving the spatial ``equation of motion''
\ba
\left[\partial_r^2+\frac{2}{r}\partial_r\right]\sigma_\omega(r)=\frac{dV_\omega(\sigma)}{d\sigma},
\ea
where 
\ba
\label{eq:potomega}
V_\omega(\sigma) = V(\sigma)-\frac{\omega}{2}\sigma_\omega^2(r),
\ea
and $V(\sigma)$ is the classical potential.
This simply follows from inserting the ansatz, Eq.~(\ref{eq:ansatz}), into the action 
\ba
S = -\int d^3x\, dt\left[(\partial_\mu\phi)^\dagger(\partial^\mu\phi)-V(\phi)\right].
\ea
\begin{figure}
\begin{center}
\includegraphics[width=8cm,angle=0,clip]{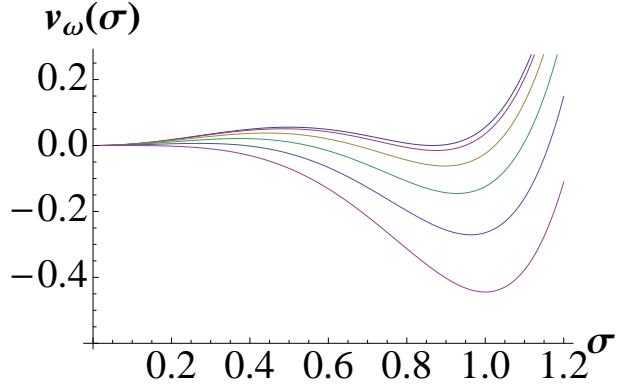}
\end{center}
\caption{The 
 effective potential $V_\omega$, for different $\omega=0$, $0.2$,
 $0.4$, $0.6$, $0.8$ and $1.0$.  $\omega=0$ corresponds to a degenerate potential, and increasing $\omega$ lowers the non-zero minimum. }
\label{fig:claspot}
\end{figure}
For the purposes of this paper we will take the potential to be (see, for instance, Ref.~\cite{mitsuo})
\ba
\label{eq:pot}
V(\phi)=m^2\phi^\dagger\phi-\lambda\left(\phi^\dagger\phi\right)^2+\frac{4g}{3}\left(\phi^\dagger\phi\right)^3.
\ea
Writing the complex field as $\phi =
(\phi_1+i\phi_2)/\sqrt{2}$, with $\phi_{1,2}$
real-valued, the potential has $\mathrm{O}(2)$ symmetry in field
space.  Let us specialise to the case where the classical potential
has degenerate minima for $v_0=0$ and $v \neq 0$ (up to global $\mathrm{O}(2)$ transformations)
\ba
\label{eq:params}
m=1,\quad \lambda = 16/3,\quad g=16/3\quad \Rightarrow \quad v=\frac{\sqrt{3}}{2}.
\ea

For $\omega=0$, the profile function $\sigma_0(r)$ can be easily found
using Eq.~(\ref{eq:pot}) directly. For non-zero $\omega$, by inserting the
ansatz as in Eq.~(\ref{eq:potomega}), we should solve for the radial
profile in the effective potential\footnote{We use ``effective'' to
  denote the potential when inserting the Q-ball ansatz, and not for
  the effective quantum potential, for which we use the term
  ``quantum''. }
\ba
V_\omega(\sigma)=\frac{m^2-\omega^2}{2}\sigma^2-\frac{\lambda}{4}\sigma^4+\frac{g}{6}\sigma^6.
\ea
Fig. \ref{fig:claspot} shows the potential for a number of different $\omega$ between 0 and 1. We now have
\ba
v^2_{\pm}=\frac{\lambda}{2g}\left(1+\sqrt{1-\frac{4}{\lambda^2}(m^2-\omega^2)}\right),
\ea
and when $\omega>m$, $\phi=0$ becomes a maximum rather than a
minimum. There is then no longer a bump separating $v_\pm$ from
$\phi=0$. Hence for some value $\omega_+\leq m$, a stable profile
function no longer exists. In general there is also a lower limit
$\omega_-$ following from an additional criterion on the potential,
$\mathrm{min}\left[V_\omega(\sigma)\right] \leq 0$, that ensures
Q-ball solutions are localised. For our case with degenerate vacua, $\omega_-=0$ \cite{mitsuo}. 

Further, in order for the Q-ball to be stable to small perturbations, the profile function must be a (local) minimum in field configuration space,
\ba
\frac{d^2S_\omega}{d\omega^2}\geq 0,\quad S_\omega=\int d^3x \left[\frac{1}{2}\left(\nabla\sigma_\omega\right)^2+V_\omega(\sigma_\omega)\right].
\ea
This concludes our brief review of classical Q-balls. An extensive literature on the subject, including many models, can be found in Ref.~\cite{Qballsrev} and references therein.

\subsection{Quantum decay: prelude}
\label{sec:quantum}

\begin{figure}
\begin{center}
\includegraphics[width=6cm,angle=0,clip]{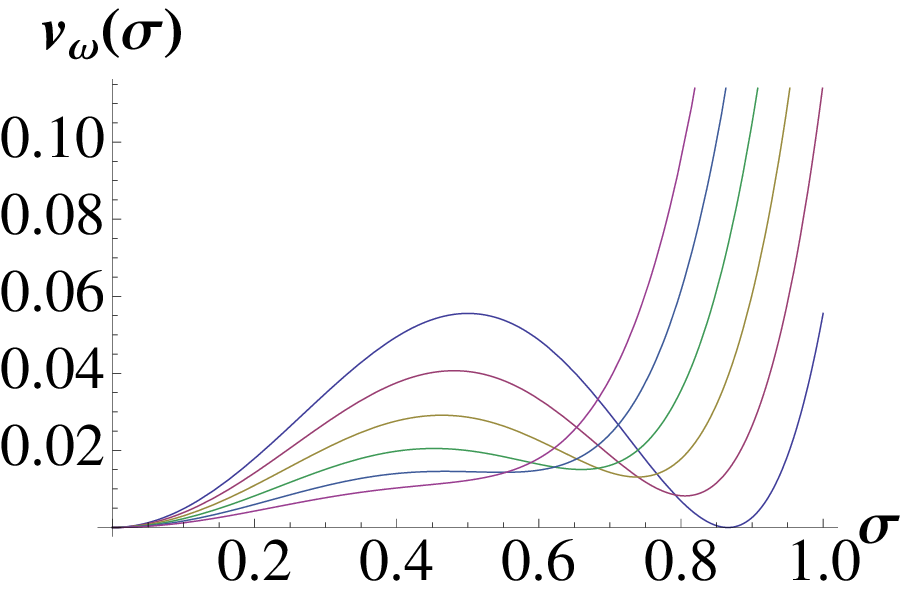}
\includegraphics[width=6cm,angle=0,clip]{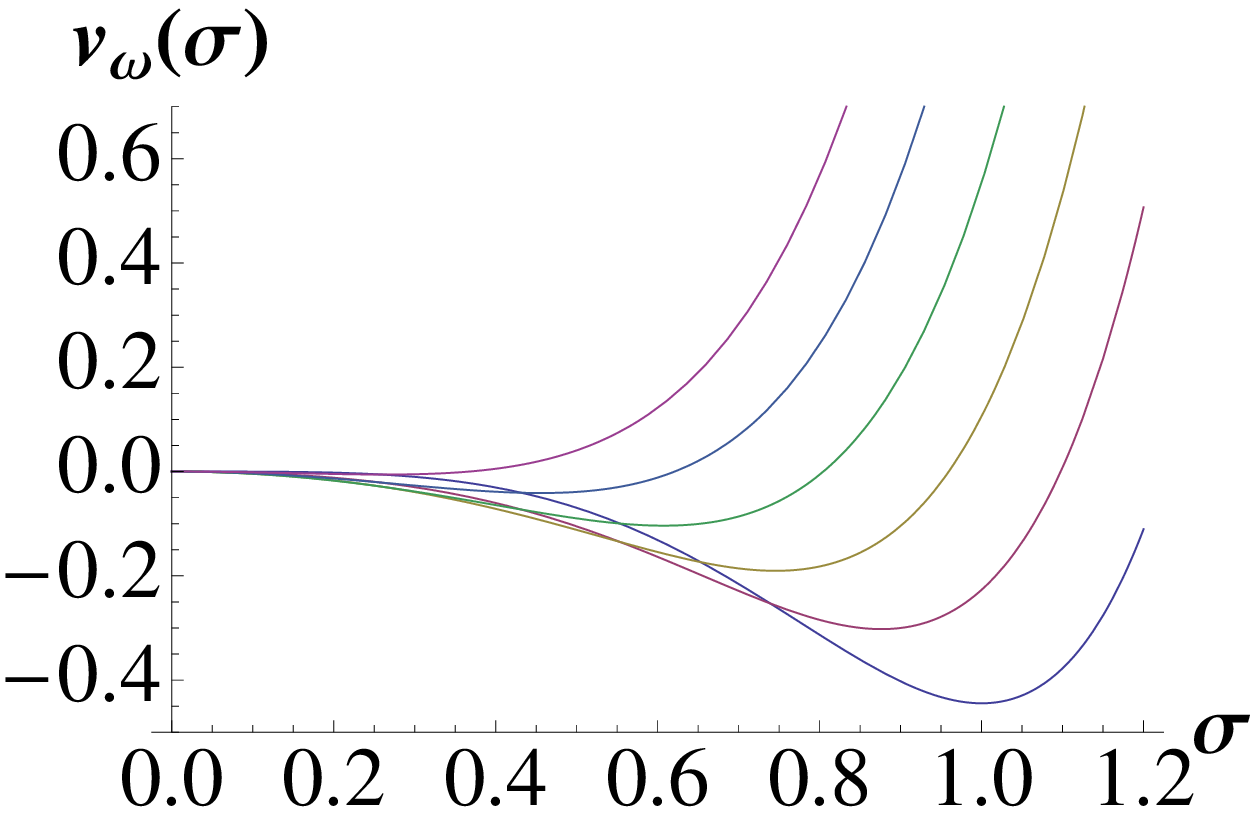}
\end{center}
\caption{The quantum effective potential $V_{\omega,\rm eff}$, for
  different values of $G$, at $\omega=0$ (left) and $\omega=1$
  (right).  Increasing $G$ lifts the potential at the nonzero
  minimum away from zero, until it is no longer a minimum. }
\label{fig:claspot2}
\end{figure}

In Ref.~\cite{mitsuo}, a detailed study of quantum stability was
carried out for the potential given in Eq.~(\ref{eq:pot})
(as well as others), in a varying number of spatial dimensions. After finding the profile functions numerically,  two quantum instabilities were considered in addition to the classical stability criterion: the possible decay of a Q-ball into a set of smaller Q-balls (with smaller Q or a set of positive and negative Q-balls adding up to the original Q); or to a set of charged, but particle-like excitations of the field. 

If the Q-ball is classically stable, decaying in this way is a tunnelling event, whereby the original profile changes into a different field configuration with particles or several Q-balls. At least in a purely quantum environment (where transitions are not mediated by thermal fluctuations) the transition rate is typically extremely suppressed. Therefore, although unstable to decay in principle, in general the Q-ball is likely to be effectively stable on the timescale of the system, $m^{-1}$. 

In the present work we seek to build upon the results of
Ref.~\cite{mitsuo}, by considering the effect of fluctuations in
changing the quantum potential in which the Q-ball evolves. Taking
the Q-ball to be the mean field (one-point correlator) of the quantum
field -- evolving in the background of quantum
fluctuations -- the
potential is modified substantially. This will alter the profile
function $\sigma_\omega(r)$ for a given $\omega$, and
hence change the range of $\omega$ within
which the criteria for stability are fulfilled.

We will perform fully real-time simulations of Q-balls and
fluctuations in the ``inhomogeneous Hartree'' approximation (see for
instance Refs.~\cite{salle1,salle2}). This will allow us to capture the
leading order effects in a quantum loop expansion, taking into account the
inhomogeneity and time-dependence of the system. Such an approach has
previously been successful when applied to topological
solitons~\cite{Bergner:2003au,Salle:2003ju,Borsanyi:2007wm}.

For such solitons, where the stability is
guaranteed by topology and the solution is time-independent, one may
do a complete Monte Carlo study as in Ref.~\cite{ArttuDavid}. This
approach completely bypasses any notion of a classical solution, which
is not possible for a time-dependent system such as a Q-ball.

\subsection{Quantum decay: homogeneous Hartree approximation}
\label{sec:homhar}

Before we embark on the technical description of this
method, we will illustrate the mechanism of instability by first considering what
happens to the potential in the ``homogeneous Hartree''
approximation. This amounts to keeping the equations of motion for the
mean field and quantum two-point functions, and throwing away all higher order
correlations. In our case, we have the connected equal time
propagators
\begin{equation}
G_1=\langle\phi_1^2\rangle - \langle\phi_1\rangle^2, \quad
G_2=\langle\phi_2^2\rangle - \langle\phi_2\rangle^2 \quad
\text{and} \quad K=\langle \phi_1\phi_2\rangle -
\langle\phi_1\rangle\langle\phi_2\rangle=\langle \phi_2\phi_1\rangle -
\langle\phi_2\rangle\langle\phi_1\rangle.
\end{equation}
In addition, $\Phi_{1,2}$ are the one-point functions of $\phi_{1,2}$.
We will derive the equations of motion in Section~\ref{sec:inhom} --
specifically, Eqs.~(\ref{eq:eom}) and~(\ref{eq:Geom}) -- but for
illustration and to motivate the approach taken in this paper we will
make use of them here. For the purposes of this section only, we make
the simplistic assumption that $G_1=G_2=G$ and $K=0$, and that they
are space and time independent. For the quantum effective potential,
this yields\footnote{Compare with Eqs.~(\ref{eq:effpotmixed}) and (\ref{eq:effpotcorrels}),
    which give the full quantum effective potential without these
    additional simplifying assumptions.}
\ba
\label{eq:Veff}
V_{\omega,\rm eff}(\sigma)&=&
\frac{m^2-\omega^2-4\,\lambda G+24\,g\,G^2}{2}\sigma^2-\frac{\left(\lambda-12\, g\, G\right)}{4}\sigma^4+\frac{g}{6}\sigma^6.
\ea
In Fig. \ref{fig:claspot2} we show the quantum effective potential for
different values of $G$ for $\omega=0$ (left hand plot) and $\omega=1$
(right hand plot). We see that the main effect of including $G$ is that it lifts and eventually removes the non-zero minimum. It is clear that when this minimum is gone, there can be no more stable Q-balls. When $\omega=1$, the non-zero minimum is the only one until for $G>\frac{1}{6}$, the zero minimum becomes the only one. 

\begin{figure}
\begin{center}
\includegraphics[width=8cm,angle=0,clip]{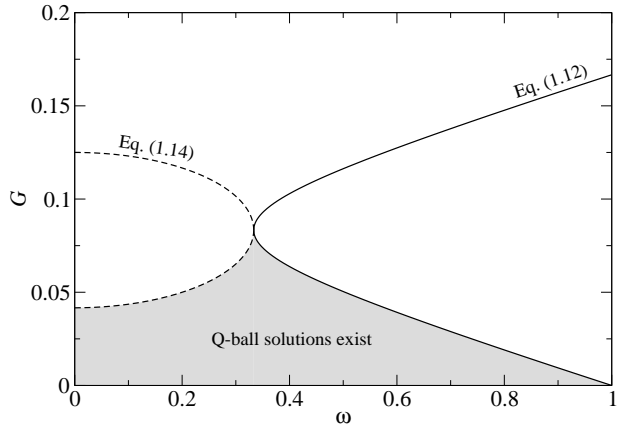}
\end{center}
\caption{Regions of two, one and no minima in the quantum effective potential. In order for the quantum potential to have two minima, $(\omega,G)$  have to be in the bottom left region, below all the curves. In particular, at $G=0$, $\omega<1$.}
\label{fig:clas_lim}
\end{figure}

The criteria for there to be two minima are that the zero-minimum
$v_0$ should have positive curvature. With our parameters, given
in~Eq.~(\ref{eq:params}), this means
\ba
\left(\frac{d^2V_{\omega,\rm eff}(\sigma)}{d\sigma^2}\right)_{\sigma=0}= (1-\omega^2)-\frac{64}{3}G+128 G^2>0,
\ea
and the non-zero minimum $v$ should exist (it always has positive
curvature). Existence of this second mimimum is when
\ba
v^2=\frac{2 - 24 G + \sqrt{1 - 32 G + 192 G^2 + 3 \omega^2}}{4}
\ea
is real and positive. When $1 - 32 G + 192 G^2 + 3 \omega^2<0$, the
non-zero minimum coalesces with the intermediate maximum that
  separates $v_0$ and $v$. And when $v^2<0$, the intermediate
maximum coalesces with the minimum $v_0$ at zero. Hence our second criterion is
\ba
(1+3\omega^2)-32 G+192 G^2>0.
\ea
We show in Fig. \ref{fig:clas_lim} how these criteria split up the $\omega$/$G$-plane into a Q-ball region (below the curves), and a single minimum region (the rest). Having $G$ and $\omega$ below the curves is a necessary criterion for the existence of Q-balls; but it is not sufficient, since then a proper profile solution has to be found which is stable, further restricting the parameter space. We see that in the classical $G=0$ limit, Q-balls are stable until some large $\omega$ close to $m=1$. But in the quantum case when $G$ is non-zero, an additional instability region opens up at low $\omega$. This is similar to the emergence of a non-zero $\omega_-$, but to distinguish it from a classical $\omega_-$, we will denote it $\omega_{\rm limit}$.

This qualitative argument also suggests that if the correlator $G$ becomes very large, stability is lost, for this particular set of renormalised parameters in the potential. Different values of the parameters of the potential will lead to modified bounds, but the picture remains the same. Also, when stability is lost in the low-$\omega$ region, it is the large-field value minimum that disappears, whereas the classical instability is the high-$\omega$ region, when the zero-field minimum becomes a maximum instead of a minimum. The latter instability also happens in the quantum case, but (potentially, depending on $G$) at smaller or larger values of $\omega$.

\subsection{This paper}
\label{sec:thispaper}

The effective quantum potential as written in
Eq.~(\ref{eq:Veff}) is unsatisfactory in a number of
ways. First of all, the correlator $K$ is nonzero in general, and
$G_{1,2}$ and $K$ are all time-dependent, since they evolve in the
background of a time-dependent mean field. In addition, they
are all space-dependent, just like the mean field. They are approximately spherically symmetric, although we will not impose that explicitly here. Also, $G_{1,2}$ are divergent quantities, and we must renormalise the equations of motion appropriately. Although the $\phi^6$ interactions formally make the theory non-renormalisable, at the level of the Hartree approximation, this can be done.

We end up solving a set of coupled differential equations for the mean
fields $\Phi_{1,2}({\bf x},t)$ and the corresponding correlators $G_{1,2}({\bf
  x,y},t,t')$. In addition, we have the cross-correlators
\begin{subequations}
\begin{align}
K({\bf x,y},t,t')&=\langle\phi_1({\bf
    x},t)\phi_2({\bf y},t')\rangle -
\langle\phi_1({\bf x},t)\rangle\langle \phi_2({\bf y},t')\rangle \\
\text{and} \quad \bar{K}({\bf
  x,y},t,t')&=\langle\phi_2({\bf y},t')\phi_1({\bf x},t)\rangle -
\langle\phi_2({\bf y},t')\rangle\langle \phi_1({\bf x},t)\rangle.
\end{align}
\end{subequations}
Note that $K$ and $\bar{K}$ coincide at equal time and space, but not
in general. In fact, we will instead make use of the method of Refs.~\cite{salle1,salle2}, where each quantum mode of the fluctuations is solved for in the background of the mean field. For reasons of computer time, we will further replace that by a classical-statistical ensemble computation of the mode correlators $G_{1,2}$ and $K$, $\bar{K}$~\cite{Borsanyi:2007wm}.

In Section 2, we will introduce the inhomogeneous Hartree approximation and the ensemble method and renormalisation of the equations of motion. Readers with less interest in the technicalities may wish to skip to Section \ref{sec:results} where we show our results for the stability region as a function of $\omega$ (note that $G$ is then no longer a free parameter, but follows from the dynamics).
We conclude in Section \ref{sec:conc}.

\section{The inhomogeneous Hartree approximation}
\label{sec:inhom}

As mentioned, we consider a self-interacting complex scalar field $\phi=(\phi_1+i\phi_2)/\sqrt{2}$, with classical action
\ba
S=-\int d^3x\, dt\,\bigg[
(\partial_\mu\phi)^\dagger(\partial^\mu\phi)+m^2\phi^\dagger\phi+\lambda(\phi^\dagger\phi)^2+\frac{4g}{3}(\phi^\dagger\phi)^3
\bigg].
\ea
We find it advantageous to recast the equations in terms of the two real-valued fields $\phi_1$ and $\phi_2$, and define the quantum mechanical one-point functions
\ba
\langle\phi_1\rangle=\Phi_1,\quad \langle\phi_2\rangle=\Phi_2,
\ea
as well as a matrix of connected two-point functions
\ba
\mathcal{G}=
\left(\begin{array}{cc}G_1(x,y)&K(x,y)\\\bar{K}(x,y)&G_2(x,y)\end{array}\right),
\ea
where
\begin{subequations}
\begin{align}
G_1(x,y) & = \langle \phi_1(x) \phi_1(y) \rangle - \langle \phi_1(x)
\rangle \langle \phi_1(y) \rangle, \\
G_2(x,y) & = \langle \phi_2(x) \phi_2(y) \rangle - \langle \phi_2(x)
\rangle \langle \phi_2(y) \rangle, \\
K(x,y) & = \langle \phi_1(x) \phi_2(y) \rangle - \langle \phi_1(x)
\rangle \langle \phi_2(y) \rangle \\
\text{and} \quad \bar{K}(x,y) & = \langle \phi_2(x) \phi_1(y) \rangle - \langle \phi_2(x)
\rangle \langle \phi_1(y) \rangle.
\end{align}
\end{subequations}
We note that
\ba
G_1(x,y)=G_1(y,x),\quad G_2(x,y)=G_2(y,x),\quad K(y,x)=\bar{K}(x,y)\;\Rightarrow\; K(x,x)=\bar{K}(x,x).\nonumber\\
\ea
We will denote $G_{1,2}(x,x)=G_{1,2}$, and similarly $K(x,x) =
  K = \bar{K}$.
The Heisenberg equations of motion read
\begin{subequations}
\label{eq:eom}
\ba
\left[\partial_t^2-\partial_x^2+m^2-\lambda(\phi_1^2+\phi_2^2)+g(\phi_1^2+\phi_2^2)^2\right]\phi_1&=&0\\
\text{and} \quad \left[\partial_t^2-\partial_x^2+m^2-\lambda(\phi_1^2+\phi_2^2)+g(\phi_1^2+\phi_2^2)^2\right]\phi_2&=&0.
\ea
\end{subequations}
We then take expectation values of Eq.~(\ref{eq:eom}), to get the equation
of motion for the mean field $\Phi_1$ (and by symmetry for $\Phi_2$),
keeping only one- and two-point functions:
\begin{subequations}
\ba
\left[\partial_\mu\partial^\mu-M_{11}^2(x)\right]\Phi_1(x)-M^2_{12}(x)\Phi_2(x)&=&0\\
\text{and} \quad \left[\partial_\mu\partial^\mu-M_{22}^2(x)\right]\Phi_2(x)-M_{21}^2(x)\Phi_1(x)&=&0
\ea
\end{subequations}
with
\begin{subequations}
\ba
M_{11}^2&=&m^2
-\lambda\left(\Phi_1^2+\Phi_2^2+3G_1+G_2\right)
+g\left(\Phi_1^4+\Phi_2^4+2\Phi_1^2\Phi_2^2\right)\nonumber\\&&
+g\left(15G_1^2+3G_2^2+10G_1\Phi_1^2+6G_2\Phi_2^2+6G_1\Phi_2^2+2G_2\Phi_1^2+6G_1G_2 \right)\nonumber\\&&
+12gK(K+\Phi_1\Phi_2)\\
\text{and} \quad M_{12}^2&=&-2\lambda K+4g K\left(\Phi_2^2+3G_1+3G_2\right);
\ea
\end{subequations}
$M_{22}^2$ and $M_{21}^2$ are obtained by interchanging the field
subscripts $1$ and $2$ in the expressions for $M_{11}^2$ and
$M_{12}^2$ respectively. We can also multiply Eq.~(\ref{eq:eom}) by
$\phi^*(y)$ from the right and take the expectation value, again
keeping only one- and two-point functions, to get
\begin{subequations}
\label{eq:Geom}
\ba
\left[\partial^2-\tilde{M}_{11}^2 (x) \right]G_1(x,y)-\tilde{M}_{12}^2
(x)
\bar{K}(x,y)&=&0,\\
\left[\partial^2-\tilde{M}_{22}^2 (x) \right]G_2(x,y)-\tilde{M}_{21}^2
(x)
K(x,y)&=&0,\\
\left[\partial^2-\tilde{M}_{11}^2 (x) \right]K(x,y)-\tilde{M}_{12}^2
(x)
G_2(x,y)&=&0,\\
\left[\partial^2-\tilde{M}_{22}^2 (x)
  \right]\bar{K}(x,y)-\tilde{M}_{21}^2 (x) G_1(x,y)&=&0,
\ea
\end{subequations}
with
\begin{subequations}
\ba
\tilde{M}_{11}^2&=&m^2-\lambda\left(
3\Phi_1^2+\Phi_2^2+3G_1+G_2
\right)
+g\left(5\Phi_1^4+\Phi_2^4+6\Phi_1^2\Phi_2^2\right)\nonumber\\&&
+g\left(15G_1^2+3G_2^2+30G_1\Phi_1^2+6G_2\Phi_2^2+6G_1\Phi_2^2+6G_2\Phi_1^2+6G_1G_2 \right)\nonumber\\&&
+12gK(K+2\Phi_1\Phi_2),\\
\text{and} \quad \tilde{M}_{12}^2&=&-2\lambda\left(K+\Phi_1\Phi_2\right) \nonumber\\&&
+4g\left(
\Phi_1^3\Phi_2+\Phi_2^3\Phi_1+3\Phi_1\Phi_2(G_1+G_2)\right)\nonumber \\
&& +12gK\left(\Phi_1^2+\Phi_2^2+G_1+G_2\right);
\ea
\end{subequations}
$\tilde{M}_{22}^2$ and $\tilde{M}_{21}^2$ being obtained as
  before.
Truncating the Schwinger-Dyson hierarchy in this way at the level of
one- and two-point functions constitutes the Hartree approximation. It
is the leading order truncation of a 2PI loop expansion (see for instance Ref.~\cite{2PIrev}).

The equations of motion are discretised in a straightforward way on a three-dimensional lattice. The differential equations are solved using a simple leapfrog algorithm in time.

In the homogeneous Hartree case, there is translational
  invariance $\Phi_1(x) = \Phi_1(t)$, $\Phi_2(x) = \Phi_2(t)$, and
  we could write
\begin{equation}
G_1(x,y)=G_1(x-y),\,
G_2(x,y)=G_2(x-y), \, K(x,y)=K(x-y) \; \text{and} \; \bar{K}(x,y)=\bar{K}(x-y).
\end{equation}
The resulting equations can be
solved numerically in a very efficient
  manner~\cite{GertHartree}. In the limit of the time dependence also being trivial, the system reduces to a simple gap equation.
However, since the Q-ball is inhomogeneous and time-dependent, these
simplifications are not possible, and we will have to solve for the
whole inhomogeneous Hartree approximation \cite{salle1,salle2}. This
means that on a $d$-dimensional lattice
with $N$ sites in each direction, the problem scales as $N^{2d}$. This is numerically possible in one spatial dimension, but at present not reliably in three. 

\subsection{Ensemble bosons}
\label{sec:ensemble}

Rather than solving the evolution in terms of the one- and two-point
functions, we will take advantage of the fact that a Gaussian system
(free or truncated at the Hartree level) may be represented in terms
of mode functions. We write (for either of the two fields $j=1,2$)
\ba
\phi_j({\bf x},t) = \Phi_j({\bf x},t) + \int \frac{d^3k}{(2\pi)^3} \left[a_{\bf k}f^j_{\bf k}({\bf x},t)+a_{\bf k}^\dagger f^{j\,*}_{\bf k}({\bf x},t)\right].
\ea
The operators $a_{\bf k}$ are time-independent by virtue of the
Gaussian approximation, and they are the standard
creation-annihilation operators obeying the relations
\ba
[a_{\bf k},a_{\bf l}^\dagger]=\delta_{\bf k,l} \quad \text{and} \quad [a_{\bf k},a_{\bf l}]=[a_{\bf k}^\dagger,a_{\bf l}^\dagger]=0.
\ea
In particular, the occupation number $n_{\bf k}$ in some state is given by
\ba
\langle a^\dagger_{\bf k} a_{\bf k}\rangle=n_{\bf k}.
\ea
Since the $a_{\bf k}$ are time independent, the numbers $n_{\bf k}$
encode all necessary information about the initial state. The mode
functions $f^j_{\bf k}(\mathbf{x},t)$ are (complex-valued) solutions of the
two-point function equations of motion given in Eq.~(\ref{eq:Geom})
\begin{subequations}
\label{eq:modeeom}
\ba
\left[\partial^2-\tilde{M}_{11}^2(x)\right]f^1_{\bf k}(x)-\tilde{M}_{12}^2f^2_{\bf k}(x)&=0\\
\text{and}\quad\left[\partial^2-\tilde{M}_{22}^2(x)\right]f_{\bf k}^2(x)-\tilde{M}_{21}^2f^1_{\bf k}(x)&=0.
\ea
\end{subequations}
In the vacuum $\Phi_1=\Phi_2=0$, the solutions are plane waves,
\ba
\label{eq:vacf}
f^j_{\bf k}(\mathbf{x},t)=\frac{1}{\sqrt{2\omega_{\bf k}}} e^{i{\bf kx}-i\omega_{\bf k} t},\qquad \omega_{\bf k}^2=k^2+m^2,
\ea
but in a general background $\Phi_i(x)$ this is no longer the case. We
will nevertheless use Eq.~$(\ref{eq:vacf})$ as our initial condition.

Numerically, discretising space on a $N^3$ lattice, there are $N^3$
mode functions for each of the two fields in this model, and so the effort of solving for them all still scales as $N^6$. Fortunately, an alternative exists~\cite{Borsanyi:2007wm}. Instead of solving at the level of $f^j_{\bf k}(x)$, and then computing
\ba
G_j(x,x) = \int \frac{d^3k}{(2\pi)^{3}} (2n_{\bf k}+1)|f^j_{\bf k}|^2(x),
\ea
one may generate a set of random numbers $\{c_{\bf k}\}$ and $\{d_{\bf k}\}$, so that for each ${\bf k}$,
\ba
\langle c_{\bf k}^*c_{\bf k}\rangle_{\rm ensemble} = \langle d_{\bf k}^*d_{\bf k}\rangle_{\rm ensemble} = n_{\bf k},
\ea
and then construct an ensemble of  realisations ($i=1,..,M$)
\begin{subequations}
\ba
\varphi_1^i({\bf x},0) = \int \frac{d^3k}{(2\pi)^3} \left[c_{\bf k}^if^1_{\bf k}({\bf x},0)+(c_{\bf k}^i)^* f^{1\,*}_{\bf k}({\bf x},0)\right],\\
\varphi_2^i({\bf x},0)= \int \frac{d^3k}{(2\pi)^3} \left[d_{\bf k}^if^2_{\bf k}({\bf x},0)+(d_{\bf k}^i)^* f^{2\,*}_{\bf k}({\bf x},0)\right],
\ea
\end{subequations}
with $f^j_{\bf k}({\bf x},0)$ given by Eq.~(\ref{eq:vacf}) at
the initial time. Then one evolves the $M$ random realisations
$\varphi_{1,2}^i$ using Eq.~(\ref{eq:modeeom}) while
simultaneously evolving $\Phi_{1,2}(x)$. At every time step one
computes
\begin{subequations}
\ba
G_1(x,x) && =
\langle\varphi_1(x)^2\rangle_\text{ensemble} - \langle
\varphi_1(x) \rangle^2_\text{ensemble} \\
G_2(x,x) && =
\langle\varphi_2(x)^2\rangle_\text{ensemble} - \langle
\varphi_2(x) \rangle^2_\text{ensemble} \\
K(x,x) && =
\langle\varphi_1(x)\varphi_2(x)\rangle_\text{ensemble} - \langle
\varphi_1(x) \rangle_\text{ensemble} \langle\varphi_2(x) \rangle_\text{ensemble}
\ea
\end{subequations}
recalling that $K(x,x) = \bar{K}(x,x)$. We should also have
$\langle\varphi_{1,2}(x)\rangle_\text{ensemble} \approx 0$. Although not exact (as computing all the mode functions would be), for $M$ large enough a very good statistical approximation results. As long as $M<N^3$, we have gained in computer efficiency; we will typically be using $N=64$ and $M=16384$, for a speed-up of a factor of 16.

\subsection{Global and local observables}
\label{sec:observables}

As our observables for tracking the evolution of the system, we will in addition to the mean fields ($\Phi_{1,2}$) and correlators ($G_{1,2}$, $K$) themselves consider the charge in the mean field
\ba
Q_{\Phi}(t)=\int d^3x\,j^0_\Phi(\mathbf{x},t)=\int d^3x\, i \left[
(\partial_t\Phi_2(\mathbf{x},t))\Phi_1(\mathbf{x},t)-(\partial_t\Phi_1(\mathbf{x},t))\Phi_2(\mathbf{x},t)
\right];
\ea
and in the modes
\ba
Q_{G}(t)=\int d^3x\,j^0_G(\mathbf{x},t)=\int d^3x\, \langle i   \left[
(\partial_t\varphi_2(\mathbf{x},t))\varphi_1(\mathbf{x},t)-(\partial_t\varphi_1(\mathbf{x},t))\varphi_2(\mathbf{x},t)\right]\rangle_{\rm ensemble};\nonumber\\
\ea
and their sum, which should be conserved. In a
  similar manner, we also consider the energy in the mean field
\begin{multline}
E_{\Phi}=\int d^3x \; \rho_{\Phi} = \int d^3x \;
\left[\frac{1}{2}(\partial_t\Phi_1)^2+\frac{1}{2}(\partial_t\Phi_2)^2+\frac{1}{2}(\partial_x\Phi_1)^2+\frac{1}{2}(\partial_x\Phi_2)^2
  \right. \\
 \left. \vphantom{\frac{1}{2}(\partial_t\Phi_1)^2}
 + V_{\rm eff}(\Phi_1,\Phi_2,G_1,G_2,K)
\right];
\end{multline}
and in the modes
\begin{multline}
E_{G}=\int d^3x \; \rho_{G} = \int \; d^3x
\left<\left[\frac{1}{2}(\partial_t\varphi_1)^2+\frac{1}{2}(\partial_t\varphi_2)^2+\frac{1}{2}(\partial_x\varphi_1)^2+\frac{1}{2}(\partial_x\varphi_2)^2
  \right.\right. \\
\left.\left. \vphantom{\frac{1}{2}(\partial_t\varphi_1)^2}
+V_{\rm eff}(G_1,G_2,K)\right]\right>_{\rm ensemble};
\end{multline}
the sum of which should again be conserved. We have split the quantum effective potential
into two parts; one with mean field and mixed terms,
\begin{multline}
\label{eq:effpotmixed}
V_{\rm
eff}\left(\Phi_1,\Phi_2,G_1,G_2,K\right)=\frac{m^2}{2}\left(\Phi_1^2+\Phi_2^2\right)-\frac{\lambda}{4}\big(\Phi_1^4+\Phi_2^4+2\Phi_1^2\Phi_2^2+\\
2G_1\Phi_2^2+2G_2\Phi_1^2+8K\Phi_1\Phi_2+6G_1\Phi_1^2+6G_2\Phi_2^2
\big)+\\
\frac{g}{6}\big(\Phi_1^6+\Phi_2^6+3\Phi_1^2\Phi_2^4+3\Phi_2^2\Phi_1^4+15\Phi_1^4G_1+15\Phi_2^4G_2+
\\
18\Phi_1^2\Phi_2^2(G_1+G_2)+3\Phi_2^4G_1+3\Phi_1^4G_2+45 \Phi_1^2G_1^2+45
\Phi_2^2G_2^2+\\
9\Phi_1^2G_2^2+9\Phi_2^2G_1^2+24\Phi_1\Phi_2K(\Phi_1^2+\Phi_2^2)+\\
72 \Phi_1\Phi_2K(G_1+G_2)+18(\Phi_1^2+\Phi_2^2)(G_1G_2+2K^2)
\big)
,
\end{multline}
and one with only $G_{1,2}$ and $K$ contributions,
\begin{multline}
\label{eq:effpotcorrels}
V_{\rm
eff}(G_1,G_2,K)=\frac{m^2}{2}\big(G_1+G_2\big)-\frac{\lambda}{4}\big(3G_1^2+3G_2^2+2G_1G_2+4K^2\big)+\\
\frac{g}{6}\big((9G_2G_1+36K^2)(G_1+G_2)+15G_1^3+15G_2^3
\big).
\end{multline}
The sum of the two is obtained by simply taking the expectation value
of the potential and keeping only one- and two-point functions.
Note that we choose to assign the interaction energy to the mean field component. 

\subsection{Renormalisation}
\label{sec:renormalisation}

The correlators $G_1$ and $G_2$ are quadratically and logarithmically divergent, so that in the vacuum with a particular mass $M^2=m^2+\delta M^2(\Phi_1,\Phi_2,G_1,G_2,K)$, we have
\begin{multline}
G_{1,2}=\int \frac{d^3k}{{(2\pi)}^3}\frac{1}{2\sqrt{k^2+m^2+\delta
    M^2_{1,2}}} \\ =\int \frac{d^3k}{(2\pi)^3}\frac{1}{2\sqrt{k^2+m^2}}-\frac{\delta M^2_{1,2}}{2}\int \frac{d^3k}{{(2\pi)}^3}\frac{1}{2(k^2+m^2)^{3/2}}+\textrm{finite}.
\end{multline}
Defining the integrands
\begin{equation}
A \equiv \int^\Lambda
\frac{d^3k}{{(2\pi)}^3}\frac{1}{2\sqrt{k^2+m^2}} \quad \text{and} \quad
B \equiv \int^\Lambda
\frac{d^3k}{{(2\pi)}^3}\frac{1}{2(k^2+m^2)^{3/2}},
\end{equation}
the most straightforward way to renormalise is to subtract  the non-field dependent part $A$ (computed as a mode sum on the finite lattice) from the correlator everywhere in the equations of motion. This amounts to a mass and a coupling renormalisation, since it is equivalent to introducing the counterterms
\begin{subequations}
\ba
m^2\rightarrow m^2+\delta m^2_{A} &=& m^2 -4\lambda A +24 g A^2,\\
\lambda\rightarrow \lambda+\delta\lambda_{A}&=&\lambda-12gA.
\ea
\end{subequations}
This approach gets rid of all quadratic divergences, leaving some
logarithmic divergences. For the parameters we will be using,
$\lambda+\delta \lambda \simeq -0.133$, which means that without
renormalisation, there would likely be no stable Q-balls at all. This
approach to renormalisation is very similar to the one used for
instance in Ref.~\cite{AT2PI}, and is easily generalised to in principle non-renormalisable potentials. It is not, however a completely rigorous 2PI renormalisation scheme (as for instance in Ref.~\cite{2PIRenorm} for renormalisable $\phi^4$ theory). For our purpose, where we do not go to the strict continuum limit, the present renormalisation scheme will be sufficient.

\section{Results}
\label{sec:results}

\begin{figure}
\begin{center}
\includegraphics[width=8cm,angle=0,clip]{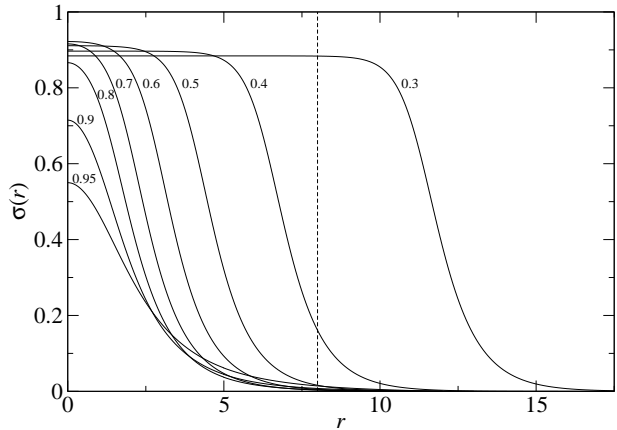}
\end{center}
\caption{The radial profiles of classical Q-balls for different values
  of $\omega$. Small $\omega$ give spatially large Q-balls. The
  vertical dashed line at $r=8$ indicates the radius of the `box'
  used to compare the charge and energy remaining within the Q-ball.}
\label{fig:profiles}
\end{figure}

We first find the classical Q-ball profile functions through a standard shooting method, very similar to Ref.~\cite{Qballsrev}. These profiles are displayed in Fig.~\ref{fig:profiles}, for a number of different values of $\omega$. Table \ref{tab:1} shows the total charge Q for a number of values of $\omega$.

\begin{table}[h!]
\begin{center}
\begin{tabular}{|c|c|c|c|c|c|c|c|c|c|}
\hline
$\omega$&0.3&0.4&0.5&0.6&0.7&0.8&0.9&0.95\\
\hline
Q&1438.5&368.5&132.5&59.4&31.3&19.1&14.4&15.0\\
\hline
\end{tabular}
\caption{Total charge for Q-balls with different $\omega$.}
\label{tab:1}
\end{center}
\end{table}

We see that small $\omega$ corresponds to large $Q$, and to large radial
size; in this limit the thin wall approximation can be
applied~\cite{Qballs}. In particular, for the size of lattices we are
able to treat numerically with the inhomogeneous Hartree
implementation at a sufficiently large mode ensemble, we were unable
to go below $\omega=0.3$. This will not influence our
findings. At the other end, large $\omega$ corresponds to small
  $Q$ and in this limit a Gaussian ansatz describes a classical Q-ball well~\cite{gleiser}.

As a check of our shooting, we evolve the initial profiles using
classical dynamics, for comparison with the Hartree evolution. Indeed,
charge and energy are well conserved
 at all times, and the Q-balls are classically
stable for all $\omega\leq 0.99$, at least on the timescale of our
simulations. In particular, we see no sign of the classical
instability reported in Ref.~\cite{mitsuo} for $\omega>0.92$. We also
checked that the results (quantum, and where applicable, classical)
were stable under variation in the number of ensemble members $M$,
lattice size and spacing and time step; the results presented here
were obtained with a lattice spacing of $a = 0.5/m$.

We define a spherical `box' around the Q-ball, for which we evaluate
the energy and charge `inside' the Q-ball, as opposed to
outside; the centre of the box tracks the barycentre of the charge in
the system. Outside energy and charge is then taken to have `decayed'
off the Q-ball. This applies to both modes and mean field separately
and combined. The box has a radius of $8$ in inverse mass
units. Further outside, at a radius $>12$ in inverse mass units, we
have a region with  damping, to minimise the amount of released energy
and charge that can reach the boundary, go around the lattice and
affect the Q-ball. This is meant to emulate the mechanism by
  which a Q-ball would decay
into the surrounding vacuum. In a simulation of many Q-balls created,
say, during a phase transition, the situation would be different. But for
our purposes here, the Q-ball is alone. We checked that the exact
choice of damping rate is not important, although it must be
non-zero. 

\begin{figure}
\begin{center}
\includegraphics[width=8cm,angle=0,clip]{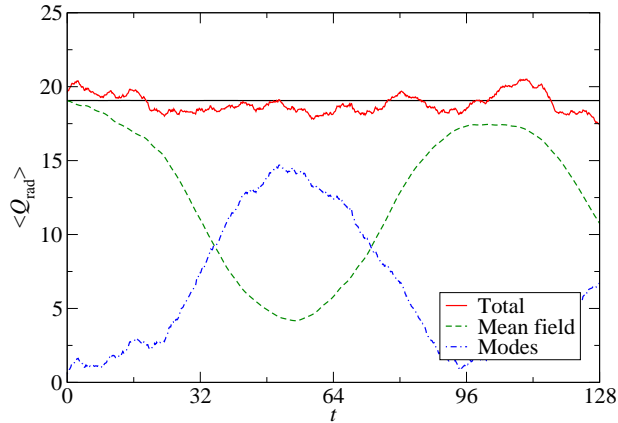}
\end{center}
\caption{The total (red; solid), mean field (green; dashed) and mode
  (blue; dot-dash) charge,
  compared to a purely classical run (black) from the same initial
  profile. Here $\omega=0.8$, and the Q-ball is stable both in
    the quantum and classical systems.}
\label{fig:charge0.8}
\end{figure}

We start with a stable example, taking $\omega=0.8$. Fig. \ref{fig:charge0.8} shows the charge inside the `box' in the
complete system (red; solid), in the mean field (green; dashed) and in the modes
(blue; dot-dash). The black dashed line is the classical simulation. We see that
the Q-ball is stable also in the quantum case, and that the charge has
the same value as for the classical case. On the face of it, this
suggests that quantum effects are very small. But we see that in the
quantum case the charge is exchanged almost completely between the
mean field and modes, an effect which simply is absent from the
classical simulation. And so although the quantum system is very
different, charge conservation is still realised as a quantum
symmetry. There is some statistical noise coming from the mode
averaging, even at $M=16384$ (it was unacceptably severe at $M=2048$), so that the charge within the
Q-ball is not quite as smooth as the classical simulation. Still, we find the agreement compelling. 

\begin{figure}
\begin{center}
\includegraphics[width=8cm,angle=0,clip]{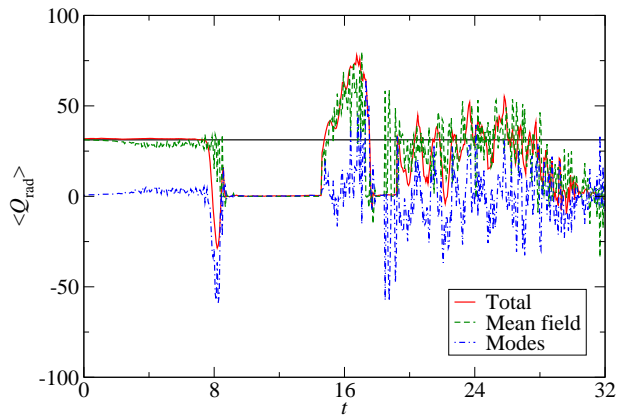}
\end{center}
\caption{The total (red; solid), mean field (green; dashed) and mode
  (blue; dot-dash) charge,
  compared to a purely classical run (black) from the same initial
  profile. Here $\omega=0.7$, and the Q-ball is unstable in the quantum system. The decay happens at around $t=8$.}
\label{fig:charge0.7}
\end{figure}

\begin{figure}
\begin{center}
\includegraphics[width=.3\textwidth,angle=0,clip]{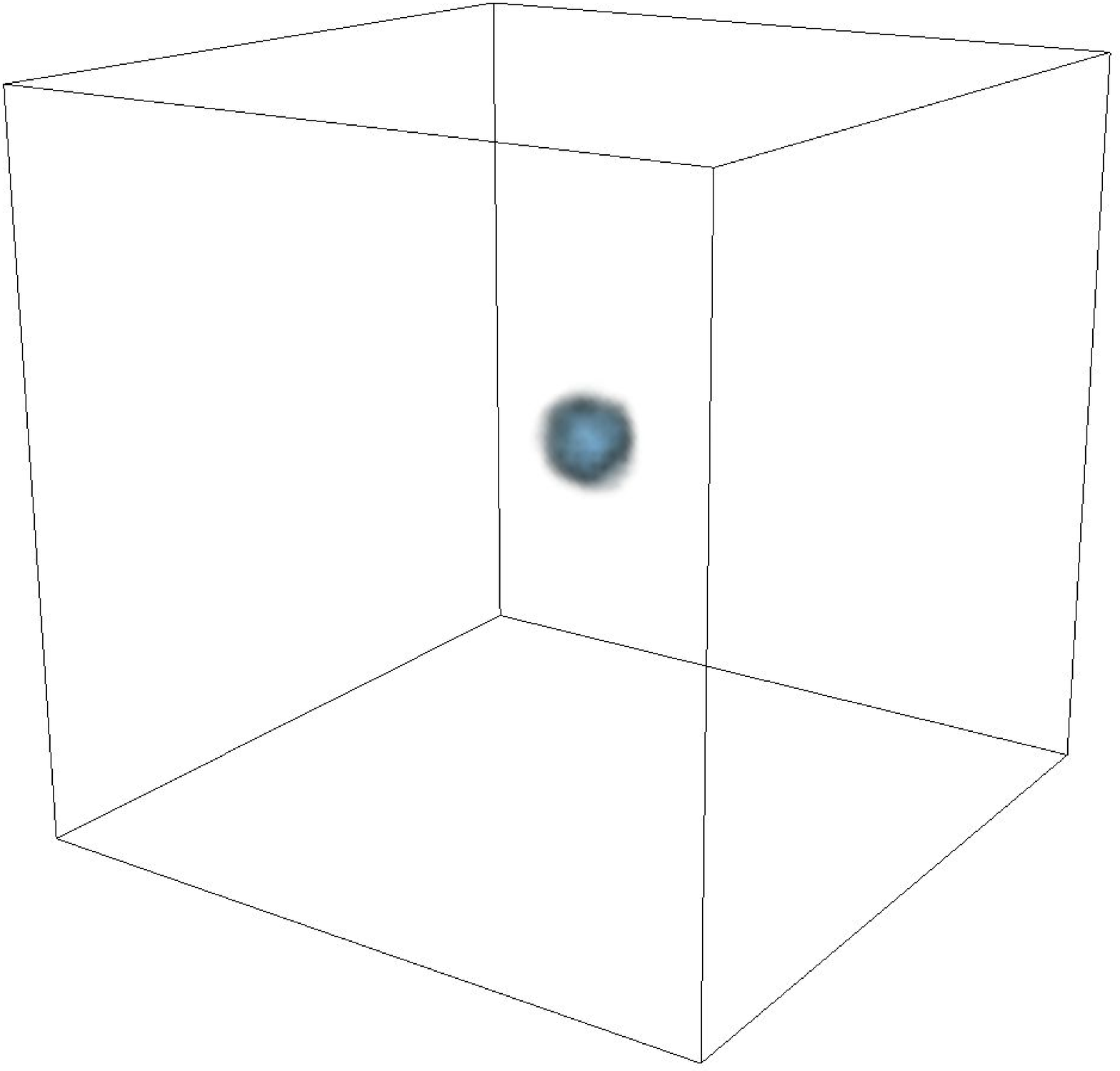}
\includegraphics[width=.3\textwidth,angle=0,clip]{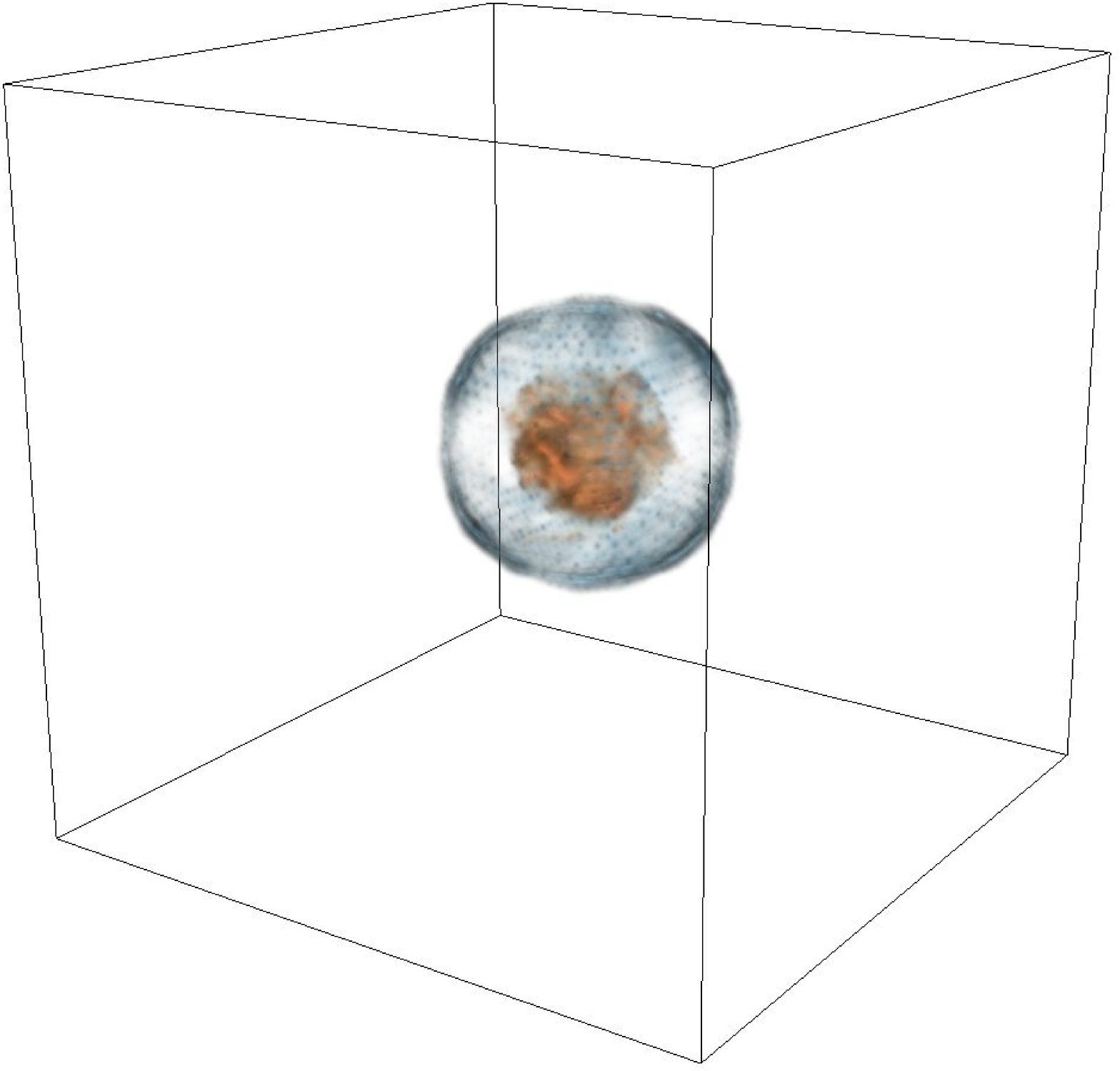}
\includegraphics[width=.3\textwidth,angle=0,clip]{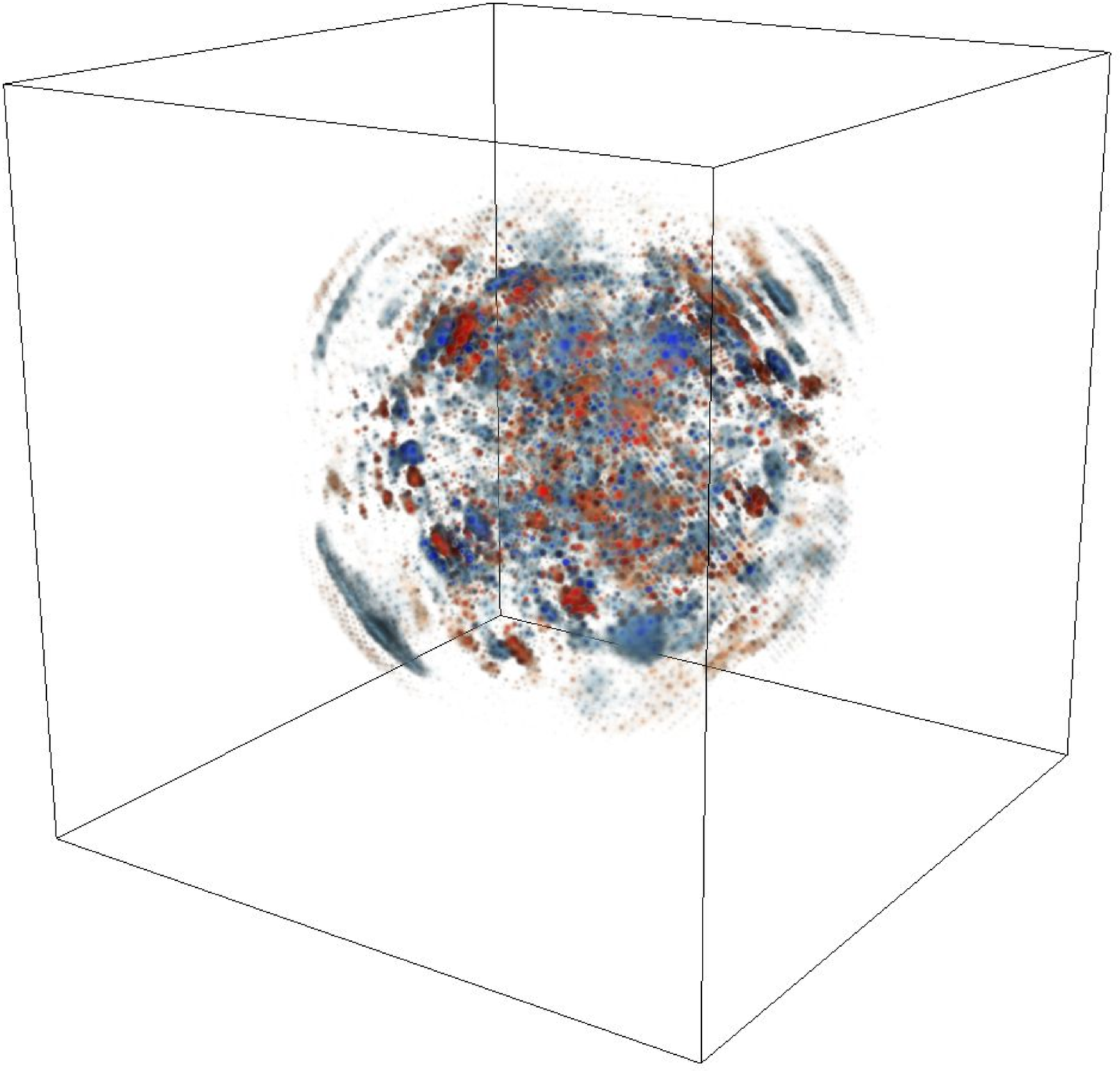}
\end{center}
\caption{Snapshots of the charge density for the $\omega=0.7$ run
  shown in Fig.~\ref{fig:charge0.7} at times $t=$ 0, 8 and 16 in inverse mass units. Blue is positive charge density, red is negative. }
\label{fig:snapshots}
\end{figure}

In Fig.~\ref{fig:charge0.7}, we show the case of $\omega=0.7$, again
inside the `box' and again the total (red; solid), mean field (green;
dashed), mode (blue; dot-dash) and classical (black) charge. The Q-ball collapses around time 8 (in inverse mass units), and after time 14 the system becomes very noisy. We see that charge completely vacates the central box volume. This is a true effect, and can be seen from considering the charge density, as shown in the snapshots in Fig.~\ref{fig:snapshots}. They correspond to times $t=$ 0, 8 and 16 in the evolution and show positive (blue) and negative (red) charge density in space.


\begin{figure}
\begin{center}
\includegraphics[width=8cm,angle=0,clip]{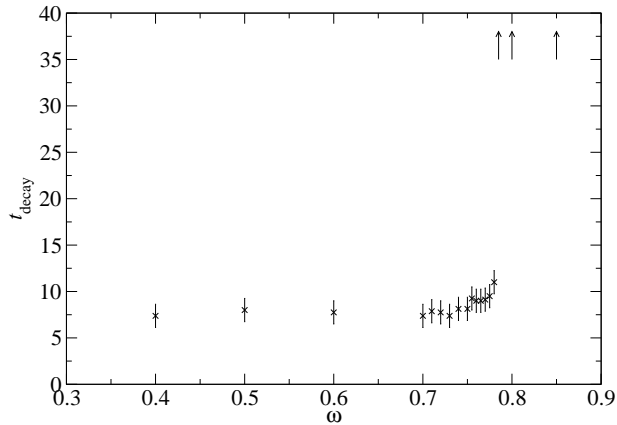}
\end{center}
\caption{The lifetime as a function of $\omega$. Simulations at and
  above $\omega=0.785$ do not decay before $t=128$, the maximum
  duration of our simulations.}
\label{fig:lifetime}
\end{figure}

We have checked that a similar collapse signature appears in the
energy inside the box. Clearly, the Q-ball is unstable in the quantum
case, but not in the classical case. In addition, we notice that in
contrast to the classical case, where instability appears above some
large $\omega$,  we now have instability below a certain limit
$\omega_{\rm limit}$. As one gets closer to this limiting value from below, the lifetime gets longer and suddenly becomes much
  longer than the duration of our simulations. Fig. \ref{fig:lifetime}
  shows these lifetimes, and we identify the limiting value as
  lying in the range
\begin{equation}
0.78 < \omega_{\rm limit} < 0.79.
\end{equation}

\begin{figure}
\begin{center}
\includegraphics[width=8cm,angle=0,clip]{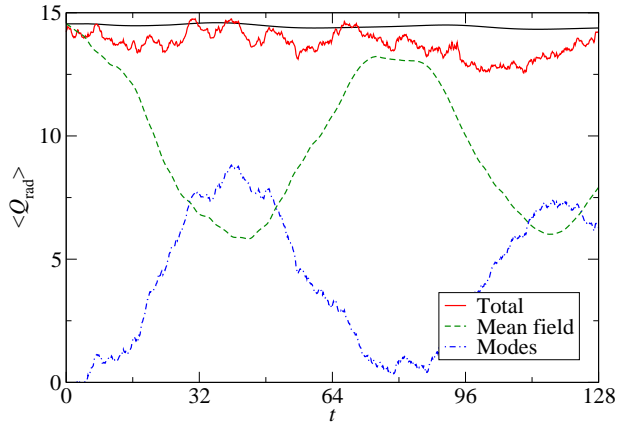}
\end{center}
\caption{The total (red; solid), mean
  field (green; dashed) and mode (blue; dot-dash) charge, compared to
  a purely classical run (black) from the same initial
  profile. $\omega=0.95$, and the Q-ball is stable also in the quantum
  system. The small oscillation in the classical charge arises
    from the damping at large radii.}
\label{fig:charge0.95}
\end{figure}

We finish by showing the charge evolution for $\omega=0.95$, which
based on the criteria of Ref.~\cite{mitsuo} is expected to be
classically unstable, and quantum unstable to fission. Our
approximation does not include fission, but nor does it explicitly
impose spherical symmetry (the initial condition is spherically
symmetric, but implemented on a cubic lattice), so collapse into
non-spherically symmetric collections of smaller objects is in
principle possible. We see in Fig.~\ref{fig:charge0.95} that the
Q-ball is both classically and quantum mechanically stable on the time
scales considered here.  Classical stability is of course lost for
$\omega \geq 1$, but even at $\omega=0.99$ we found classical
stability in the real-time numerical evolution.

\section{Conclusion}
\label{sec:conc}

Inspired by the comprehensive work of Ref.~\cite{mitsuo}, we have implemented the inhomogeneous Hartree approximation to quantum dynamics for Q-balls in a particular classical potential, with a particular set of parameters. The classical potential is of the ``degenerate minima'' type, where there are two classical minima, separated by a maximum. Based on general criteria, Q-balls are expected to be unstable to quantum decay into smaller objects for $\omega>0.82$, and otherwise completely stable. They are classically unstable for even larger $\omega>0.92$, as the minimum at $\phi=0$ becomes too shallow to support a Q-ball, and effectively disappears.

Our findings represent interesting complementary information to this
result. To leading order in a 2PI loop expansion, we found that
quantum corrections to the effective potential in which the mean field
Q-ball lives change the picture completely. There is now a
limiting $\omega$ around 0.78-0.79 {\it below} which Q-balls are
unstable to collapse. Not through tunnelling or decay as for the
quantum instability discussed in Ref.~\cite{mitsuo}, but
classical-like collapse as the non-zero minimum becomes shallow
and -- perhaps -- effectively disappears. Above this limiting
frequency, the Q-balls are stable at least up to the maximum
$\omega=0.99$ tested.

We are not able to go beyond $\omega=1$, since there are no classical
initial profiles there. In addition, for numerical reasons we are also unable to go below $\omega=0.3$ to see whether stability
reappears. As far as we have been able to go, there is no sign of this, and our qualitative analysis in Section~\ref{sec:quantum} suggests the same.

The main caveat to these conclusions is that the chosen potential has
quite large bare parameters, and considering the Hartree approximation
as a coupling expansion it is possible that this truncation is not
reliable. In a sense, the exact quantum effective potential would
include the precise local potential form, quantum tunnelling and non-perturbative decay. For an unstable mean field configuration, it may even have a physically significant imaginary part, signalling this instability. Clearly, the Hartree approximation does not include tunnelling, but is here assumed to give the dominant contribution to the local shape of the potential near the minima, ultimately changing minima and maxima into maxima and minima. One may fear that this is a crude approximation. On the other hand, the Hartree approximation is an infinite resummation of perturbative diagrams. Order counting in a 2PI expansion also involves the magnitude of the propagator, which is rather small here. Ideally, one would like to go to the next order in this expansion, but currently 2PI-NLO is not numerically tractable for an inhomogeneous system. 

In summary, when studying the
quantum stability of Q-balls it is perhaps more important
to understand the modification of the mean field potential, than the
non-perturbative decay and tunnelling of the classical
solutions. This is because tunnelling typically occurs on very long timescales, although
estimating the decay rate is very difficult. It is not clear
that the classical profile function is the correct starting point for
such a tunnelling transition, since the quantum potential is
different. Furthermore, when considering the corrected quantum
effective potential (in our case in the Hartree approximation), one
finds a qualitatively different stability pattern as a function of
$\omega$. The non-zero minimum can disappear for large $G$ and Q-ball
solutions no longer exist. This happens on a very short timescale of a
few tens of inverse mass units (essentially the time it takes for the
quantum effective potential to settle in our simulations).

We have argued that an analysis of the instability (as in Section
\ref{sec:quantum}) gives the correct qualitative understanding of the
Q-ball decay, but
that quantitative understanding would require large-scale
simulations. In particular, we found that at
  least $M=16384$ mode
realisations are necessary for statistical convergence of the
method when $N=64$, and for instance that $M=2048$ is insufficient.

Finally, we have shown that the quantum effects are significant in
this model, as charge is almost
completely transferred between the mean field and the quantum
modes. This is despite the general consensus that a Q-ball will
essentially behave classically. That assumption needs to be more rigorously
tested in future work.

Many classical simulations of Q-ball formation, evolution and
interactions have been carried
out~\cite{Battye:2000qj,Enqvist:2000cq,Multamaki:2002hv,Hiramatsu:2010dx}. The
present work demonstrates that the dynamics of quantum Q-balls may be
very different from that of classical Q-balls, and that it is
feasible to simulate quantum Q-ball behaviour. Such Q-ball simulations
should therefore be revisited.

For precision computations, it may be relevant to improve on the
renormalisation procedure, since in addition to physical modifications
of the potential, some effects may remain of using our approximate
subtraction scheme. The main result of this is that at finite lattice
spacing, matching the renormalised couplings to be equal to the
classical ones is not exact, as one would ideally wish. The effective modification of the potential is therefore in small part due to
this. This may be remedied order by order in a loop expansion,
although the fully resummed counterterm procedure of Ref.~\cite{2PIRenorm} may
not generalise to non-renormalisable potentials.

The obvious generalisation is to consider other parameter choices and different theories of Q-balls, in addition to the single example presented here. It could also be interesting to do a completely classical-statistical simulation of the system as an approximation to the quantum Q-ball system. This would involve generating an appropriate ensemble of initial conditions to evolve classically and average over. These projects are both underway. 

\vspace{0.2cm}

\noindent {\bf Acknowledgments:} The numerical simulations were
performed on the Norwegian computer cluster Abel, under the NOTUR
project. AT thanks the Villum foundation for support and DJW thanks
Mark Hindmarsh and Arttu Rajantie for stimulating discussions. We thank Paul Saffin for useful comments.

\end{document}